\documentclass[traditabstract, letter]{aa}

\usepackage{graphicx}
\usepackage{txfonts}
\usepackage{natbib}
\bibpunct{(}{)}{;}{a}{}{,}
\begin{document}

\author{
  B. Klein\inst{1}\fnmsep\inst{2}   \and
  S. Hochg\"urtel\inst{1}           \and
  I. Kr\"amer\inst{1}               \and
  A. Bell\inst{1}                   \and
  K. Meyer\inst{1}                  \and
  R. G\"usten\inst{1}
}

% \offprints{Bernd Klein}

\institute{
  Max-Planck-Institut f\"ur Radioastronomie, Auf dem H\"ugel 69, 53121 Bonn, Germany \\
  \email{bklein@MPIfR-Bonn.MPG.de}
  \and
  University of Applied Sciences Bonn-Rhein-Sieg, Grantham-Allee 20, 53757 Sankt Augustin, Germany
}

\title { High-resolution wide-band Fast Fourier Transform spectrometers }

\date{Received 22 January 2012 / Accepted 12 March 2012}

 \abstract
 {
   We describe the performance of our latest generations of sensitive wide-band high-resolution
   digital \underline{F}ast \underline{F}ourier \underline{T}ransform \underline{S}pectrometer (FFTS).
   Their design, optimized for a wide range of radio astronomical applications, is presented.
   Developed for operation with the GREAT far infrared heterodyne spectrometer on-board SOFIA,
   the eXtended bandwidth FFTS (XFFTS) offers a high instantaneous bandwidth of 2.5\,GHz with
   88.5\,kHz spectral resolution and has been in routine operation during SOFIA's Basic Science
   since July 2011. \\
   We discuss the advanced field programmable gate array (FPGA) signal processing pipeline,
   with an optimized multi-tap polyphase filter bank algorithm that provides a nearly loss-less
   time-to-frequency data conversion with significantly reduced frequency scallop and fast sidelobe
   fall-off.
   Our digital spectrometers have been proven to be extremely reliable and robust, even under the harsh
   environmental conditions of an airborne observatory, with Allan-variance stability times of several
   1000\,seconds. \\
   An enhancement of the present 2.5\,GHz XFFTS will duplicate the number of spectral channels (64$k$),
   offering spectroscopy with even better resolution during Cycle\,1 observations.
 }

\keywords{ Instrumentation: spectrographs -- Techniques: spectroscopic }
\maketitle

%
%------Start of the normal Text --------------------------------------------------------------------------
%

%%% INTRODUCTION
\section{Introduction}
\label{Introduction}

%%% FIGURE :: SPECTROMETER TYPES
\begin{figure}
  \centering
   \includegraphics[width=8.7cm]{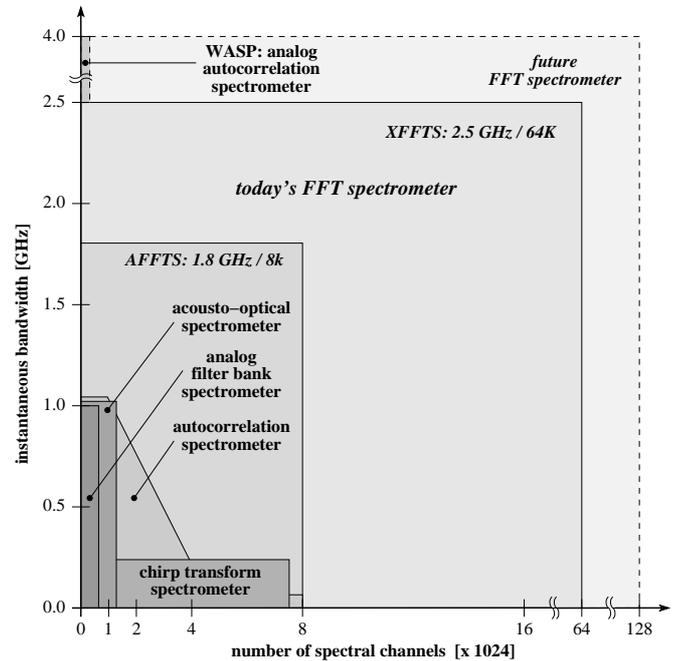}
  \caption{
     Comparison of different spectrometer types in terms of instantaneous bandwidth and spectral
     channel number applied in radio astronomy. Only FFT spectrometers provide both wide bandwidth
     and high-frequency resolution because of their impressive number of independent spectral channels
     \citep{KleinSPIE:2006}.
  }
  \label{SpectrometerTypes}
\end{figure}

Until the recent past signal processing instruments for radio-astronomical spectroscopy were highly
dedicated developments, customized to the individual scientific application (Fig.1).
Digital auto-correlators have been widely-used in the radio-astronomical communities since the
detection in 1963 of the first interstellar molecule at radio wavelength, the hydroxyl radical
OH \citep{WeinrebNature:1963}, with the auto-correlation technique \citep{Weinreb:1963}.
In a late version, the array auto-correlator serving the CHAMP heterodyne
array \citep{Wiedenhoever:1998, Kasemann:2006} used the high-performance (\emph{Canaris}) CMOS chip
for a maximum bandwidth of 1\,GHz with 1024 lags.
In parallel, acousto-optical spectrometers (AOS) have been widely used, with comparable bandwidth limitations
\citep{Cole:1968, Schieder:2003}.
For high spectral resolution applications, with limited bandwidth requirements, chirp transform spectrometers
(CTS) are in operation \citep{Hartogh:1997}.
The early back-end concept of GREAT, when defined almost 10 years ago, was consequently based on
a suite of wide-band (1\,GHz) AOS (with 1\,MHz channel spacing) and -- for high resolution applications --
CT-spectrometers (200\,MHz, 56\,kHz resolution) only.

Then, less than a decade ago, the exploding computing power of high-performance field programmable
gate array (FPGA) chips and the rapidly increasing sampling rate of commercially available analog-to-digital
converters (ADC) revolutionized the field, and \emph{Fourier transform} spectroscopy became possible.
In this process, the down-converted intermediate frequency signal of the coherent receiver
is first sampled at high resolution, then -- in the FPGA -- the Fast Fourier Transform (FFT) is
calculated and the spectra are summed (see \citealt{KleinAA:2006} for details and a visualization
of the Wiener-Khinchin theorem).

\begin{figure}
  \centering
   \includegraphics[width=8.7cm]{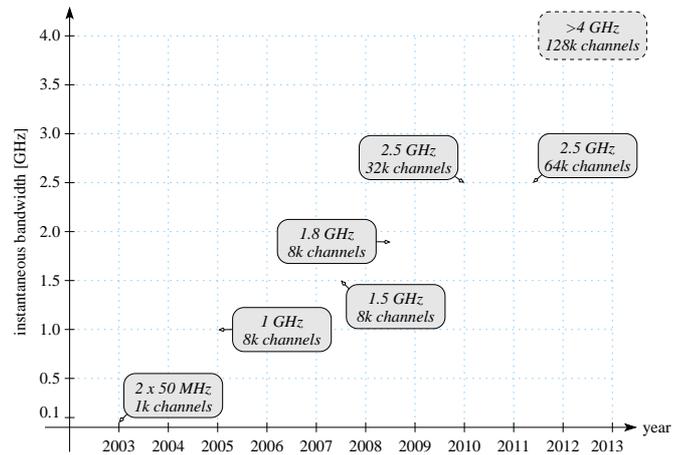}
  \caption{
     Overview of the FFTS developments at the Max-Planck-Institut f\"ur Radioastronomie.
     In the last 8 years it was possible to improve the instantaneous bandwidth by a
     factor of 50 while processing 64 times more spectral channels.
     The dashed box marks the design goals for our next generation of FFTS.
  }
  \label{FFTShistory}
\end{figure}

\cite{Stanko:2005} performed successful astronomical observations already in August 2004 at the
100m-telescope with a first fully digital Fast Fourier Transform spectrometer (though at that time
limited to just 2\,$\times$\,50\,MHz bandwidth and 1024 (1$k$) spectral channels).
Only shortly later, the rapid advances in digital signal processing (DSP) hardware made it possible
to develop the first wide-band FFT spectrometers with already 1\,GHz bandwidth \citep{Benz:2005, KleinAA:2006}.
In 2007 our Array FFTS (AFFTS) with up to 1.8\,GHz bandwidth and 8$k$\,channels saw first light 
\citep{Klein:2008}, and here we present the latest product of these amazing advances in DSP:
the  XFFT spectrometer for GREAT\footnote{GREAT is a development by the MPI f\"ur Radioastronomie
  and the KOSMA\,$/$\,Universit\"at zu K\"oln, in cooperation with the MPI f\"ur Sonnensystemforschung
  and the DLR Institut f\"ur Planetenforschung.}
now analyzes 2.5\,GHz of instantaneous bandwidth in 32$k$ spectral channels (Fig.\,\ref{FFTShistory}).

With the availability of the first samples of once again faster analog-to-digital converters and
new FPGAs with still increasing processing capabilities (Fig.\,\ref{FFTShistory}),
present-day FFT spectrometers meet all requirements of high-resolution spectroscopy in
radio astronomy and beyond, thereby replacing traditional spectrometers.
The FFTS provide wide bandwidth with very fine spectral resolution at comparatively low production costs.
The high dynamic range of today's ADCs with 8/10-bit allows observing strong continuum sources and bright
maser lines without signal loss.
The FFT spectrometers operate very stably with long Allan-variance times \citep{Klein:2008};
they are calibration- and aging free and perform extremely reliable in the harshest environments.
The generic approach of our signal processing pipeline allows generating FPGA cores with different
bandwidths and/or spectral resolution on a short time scale. \\

In this paper we describe the hardware and the FPGA signal processing of the AFFTS and XFFTS
developed for APEX\footnote{APEX is a collaboration between the Max-Planck-Institut
  f\"ur Radioastronomie, the European Southern Observatory, and the
  Onsala Space Observatory.}
and GREAT \citep{Heyminck:2012}.
First we discuss the goal and design of the Array FFT Spectrometer (Sect.\,\ref{AFFTS}),
followed by a description of our latest development -- the XFFTS -- in Section \ref{XFFTS}.
In Section \ref{FFTSsignalProcessing} we outline the FPGA signal processing and the polyphase
filter bank algorithm as well as the spectral behavior and frequency response of our processing pipeline.
Section \ref{FFTSspecification} summarizes the technical details of the two spectrometers and their
configuration and performance during now worldwide field operations (Section \ref{FFTSinstallations}).
Finally, we present an outlook on on-going and future FFTS developments (Sect.\,\ref{FFTSoutlook}). \\

%%% FIGURE :: FFTS DEVELOPMENT HISTORY

%% AFFTS -------------------------------------------------------------------------------------------
\section{AFFTS -- the Array FFT Spectrometer}
\label{AFFTS}

%%% FIGURE :: AFFTS BLOCK DIAGRAM
\begin{figure}
  \centering
   \includegraphics[width=8.7cm]{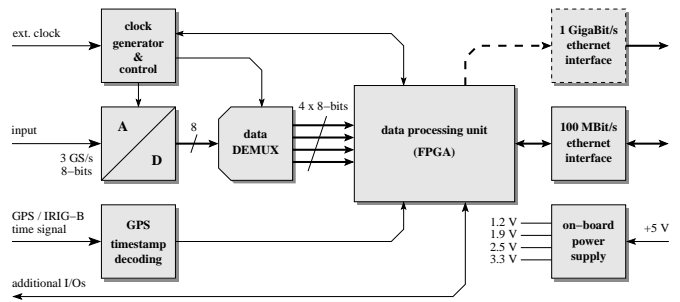}
  \caption{
     Block diagram of the 1.5\,GHz bandwidth AFFTS digitizer/analyzer board.
     The board can be equipped with a single- or dual-input ADC (National ADC083000
     or ADC08D1500). The optional GigaBit-ethernet interface allows high-speed
     data acquisition with transfer rates of up to 90\,MBytes/s, which are required
     for pulsar/transient search or to readout MKIDs -- \underline{M}icrowave
     \underline{K}inetic \underline{I}nductance \underline{D}etectors.
  }
  \label{AFFTSblockdiagram}
\end{figure}

To serve the operation of our single pixel detectors (a.o. GREAT), but in particular
to process the $\sim$\,50\,GHz signal bandwidth of mid-sized heterodyne arrays like
CHAMP$^+$/APEX \citep{Guesten:2008} or EMIR/IRAM \citep{Carter:2012}, and the future
upGREAT for SOFIA, in 2007 we launched the development of a 1.5\,GHz bandwidth
digitizer-/analyzer-board: the Array FFT Spectrometer (AFFTS).\\
The development goal was to design a compact and robust digital spectrometer board that
can be produced inexpensively with simply commercially available components.
The 100\,$\times$\,160\,mm sized AFFTS-board combines a National 3\,GS/s ADC (ADC08300)
and a Xilinx Virtex-4 SX55 FPGA.
The wide analog input bandwidth of the 8-bit ADC enables one to sample the input signals
at baseband (DC to 1.5\,GHz) or in the second Nyquist zone (1.5\,$-$\,3.0\,GHz).
The  boards include a standard 100\,MBits/s ethernet interface, which simplifies
the combination of many boards into an Array FFTS via a common ethernet switch.
A block diagram of the AFFT-board is presented in Figure\,\ref{AFFTSblockdiagram}.
The spectrometer board operates from a single 5\,Volt source and dissipates less than
20\,Watt, depending on the actual configuration of bandwidth and number of spectral
channels.
Precise time stamping of the processed spectrum is realized by an on-board GPS/IRIG-B
time decode circuit.
Furthermore, the 10-layer digital board includes a programmable ADC clock synthesizer
for a wide range of bandwidth configurations (0.1\,$-$\,1.8\,GHz), making the spectrometer
flexible for different observing requirements and receiver characteristics.\\

%% XFFTS -------------------------------------------------------------------------------------------
\section{XFFTS -- the eXtended bandwidth FFTS}
\label{XFFTS}

%%% FIGURE :: XFFTS BOARD
\begin{figure}
  \centering
   \includegraphics[width=8.7cm]{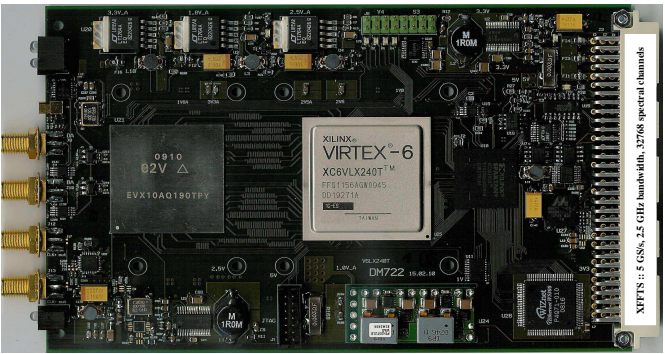}
  \caption{
     XFFTS board that uses E2V's fastest 4\,$\times$\,1.25\,GS/s 10-bit ADC
     and the high-performance Xilinx Virtex-6 LX240T FPGA.
     By applying time interleave techniques, the four 1.25\,GS/s ADCs can be
     combined into two 2.5\,GS/s or one 5\,GS/s converter.
     The XFFTS allows analyzing an instantaneous bandwidth of 2.5\,GHz with
     32$k$ spectral channels.
     Depending on the processing set-up and the input signal level, the board
     consumes 20\,$-$\,25\,Watt.
  }
  \label{XFFTboard}
\end{figure}

%%% FIGURE :: XFFTS CRATE
\begin{figure}
  \centering
   \includegraphics[width=8.7cm]{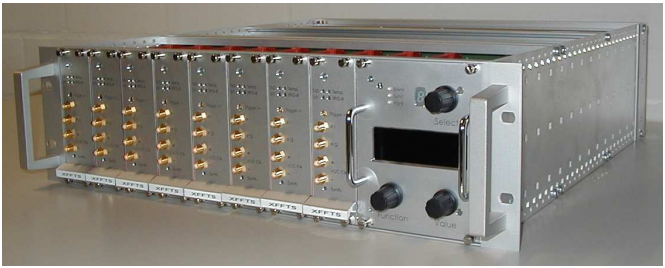}
  \caption{
     Photograph of the 19-inch XFFTS-crate, equipped with eight XFFTS-boards
     (hence processing 20\,GHz signal bandwidth) and one FFTS-controller unit.
     The modular concept allows combining multiple crates to build large
     FFT spectrometer arrays.
  }
  \label{XFFTScrate}
\end{figure}

With the availability of first samples of E2V's 5\,GS/s 10-bit ADC in mid 2009, we
commenced the development of the wide-band XFFTS board \citep{Klein:2009}.
The goal was a digitizer-analyzer-board with broader bandwidth and higher spectral
resolution than provided by the AFFT to analyze the instantaneous bandwidth of the
GREAT HEB receivers ($>$\,2\,GHz) in one monolithic part.
To realize an adequate number of spectral channels, we designed the new XFFTS board
around the high-performance Xilinx Virtex-6 LX240T FPGA.
This device is able to process a continuous data flow of 6\,GBytes/s in a polyphase
filter bank with 32768 spectral channels.

Because E2V's 5\,GS/s ADC is not a monolithic converter but a fourfold 1.25\,GS/s ADC,
the four ADCs have to be time-interleaved to synthesize the behavior of a 5\,GS/s converter.
The main challenges with time-interleaving are accurate phase alignment of
sampling-clock edges between channels, and compensation for manufacturing variations
that inherently occur between ICs.
Accurately matching the gain, offset, and clock phase between separate ADCs is very challenging,
especially because these parameters are frequency-dependent \citep{McCormack:2009}.
For the XFFTS we developed an adaptive FPGA-based calibration routine that measures
an injected fixed frequency continuous wave line and calculates the best parameter
for gain, offset, and clock phase.
Applying this optimized calibration scheme, no interleaving artifacts are noticed,
even in long integrations.

Similar to the AFFTS, the XFFTS-board includes the GPS/IRIG-B time decoder unit and a
low-jitter on-board ADC synthesizer.
In Figure\,\ref{XFFTboard} we show a photo of the 12-layer XFFTS board.
Up to eight XFFTS-boards can be housed in one 19'' FFTS-crate together with power
supplies (4\,$\times$\,5\,Volt, 20\,Amperes) and one FFTS-controller (Fig.\,\ref{XFFTScrate}).
The latter is responsible for the distribution of global synchronization signals,
e.g., the reference clock for the ADC synthesizer, or the GPS/IRIG-B timing information.
In addition, the controller displays housekeeping information such as board IP numbers,
temperatures of the ADC and FPGA chips and the power level of the IF inputs.
All controller informations are also available by ethernet, making a completely remote
operation of the XFFTS possible.  \newline
To serve the need for spectroscopy with even higher spectral resolution, we enhanced
the present XFFTS by replacing the FPGA with the substantially more complex Xilinx Virtex-6
SX315T chip to implement 64$k$ spectral channels.

%% Advanced FPGA signal processing ----------------------------------------------------------------
\section{Advanced FPGA signal processing}
\label{FFTSsignalProcessing}

%%% FIGURE :: SIGNAL PROCESSING - POLYPHASE FILTER BANK (PFB)
\begin{figure}
  \centering
   \includegraphics[width=8.7cm]{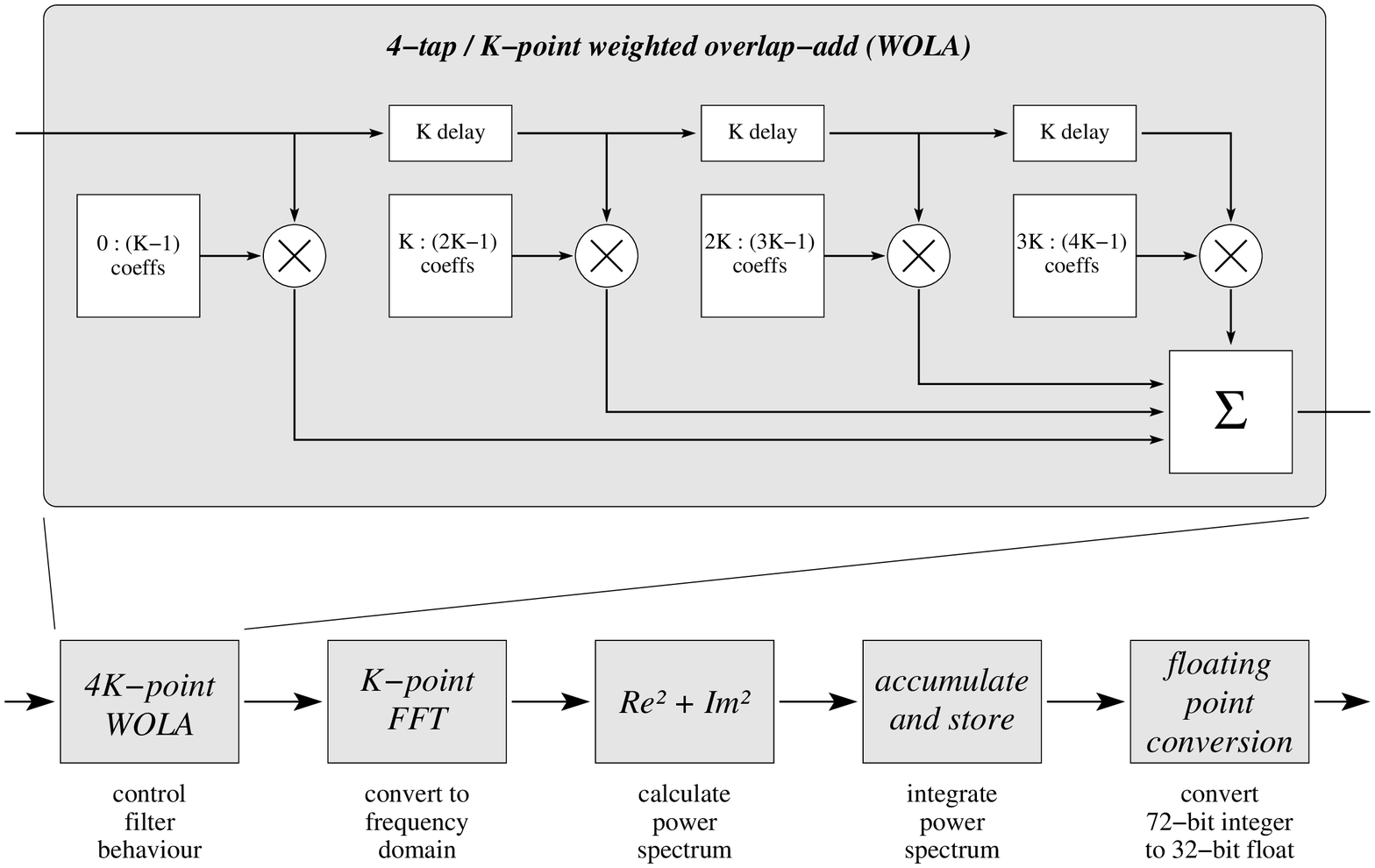}
  \caption{
     Block diagram of the polyphase filter bank (PFB) signal processing pipeline.
     The PFB not only produces a flat response across the passband, but also
     excellently suppresses of out-of-band signals.
  }
  \label{WOLAprocessing}
\end{figure}

As described in \cite{KleinSPIE:2006}, the complete signal processing pipeline for
converting time-sampled data into an integrated power spectrum fits on a single
complex FPGA chip.
The advanced spectrometer core for our FFTS-boards is an in-house development
by MPIfR and based on a generic VHDL\footnote{
 VHDL: \underline{V}irtual \underline{H}ardware \underline{D}escription \underline{L}anguage}
approach, which uses Xilinx's IP core-generator to configure the pipeline-FFTs.
This allows a maximum of flexibility in FFT-length, FFT-bit-width, and processing speed.
Unlike the commonly applied window in front of the FFT to control the frequency response,
a more efficient polyphase pre-processing algorithm has been developed with significantly
reduced frequency scallop loss, faster side lobe fall-off and less noise bandwidth
expansion.
Figure\,\ref{WOLAprocessing} illustrates the polyphase filter bank (PFB) signal processing
in the FPGA: After the polyphase filter that we implemented as a pipeline version of the
weighted overlapp-add (WOLA) method \citep{Crochiere:1983}, the FFT is realized using a
highly parallel architecture to achieve the very high data rates of 3\,GBytes/s (AFFTS) and
6\,GBytes/s (XFFTS).
The next step of the signal processing includes converting the complex frequency spectrum
to a power density representation and successive accumulation of these results by an
adjustable time-period.
This accumulation step has the effect of averaging a number of power spectra,
thereby reducing the background noise and improving the detection of weak signal
parts.
In addition, this step also reduces the huge amount of data produced by the prior
processing stages and eases any subsequent interfacing for the data analysis.

%%% FIGURE :: 32K CHANNEL SPECTROMETER DESIGN
\begin{figure}
  \centering
   \includegraphics[width=8.7cm]{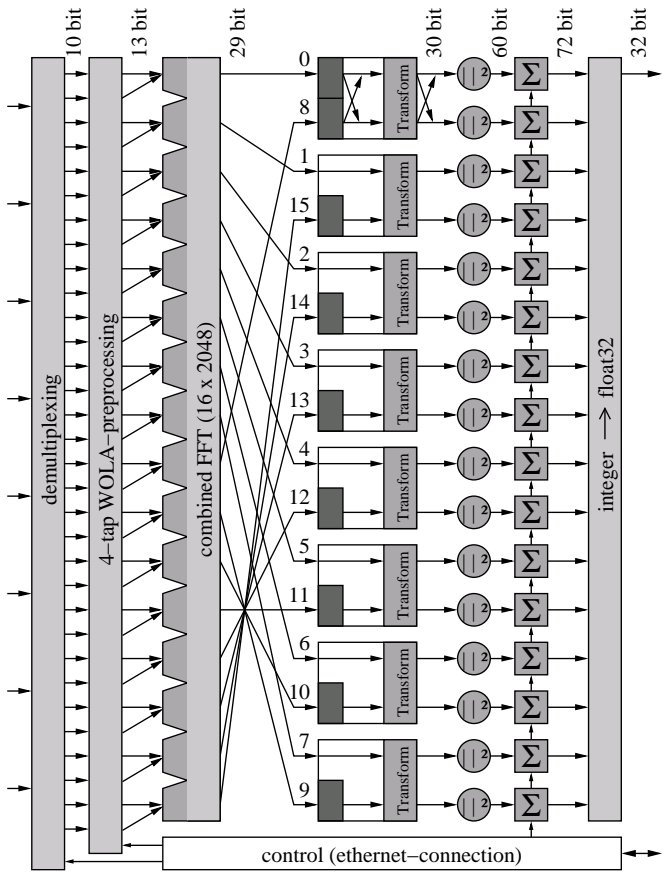}
  \caption{
     32$k$ spectrometer design with a combined 32768-points FFT (16 $\times$ 2048),
     running on a Xilinx\,Virtex6 LX240T FPGA.
     Data width grows by $log(4)+1$ bits in four-tap WOLA,
     by $log(32768)+1$ bits in 32768-points FFT,
     by $1$ bit during channel transformation and is doubled by squaring.
  }
  \label{32kSpectrometerCore}
\end{figure}

All signal processing steps use unrounded computations to prevent a loss of
sensitivity during the calculations.
The data bandwidth after each processing unit is equal to the data bandwidth
before plus the logarithm of the number of summed values:
$log(4)$ bits in the four-tap WOLA and $log(N)$ in the N-point FFT and channel-transformation
unit. Two extra bits are spent to compensate for window-multiplication and to prevent an
overflow by the twiddle-multiplication in the FFT.
The
% $I^2 + Q^2$
power builder doubles the data width and up to 72\,bits are finally reserved
for each channel of the integrated power spectrum.
Last step in the processing pipeline is the conversion from integer representation
(AFFTS: 64-bit, XFFTS: 72-bit) to 32-bit floating-point format.
The spectrometer design of the 32$k$ channels XFFTS core with the bit-width in each stage
is illustrated in Figure\,\ref{32kSpectrometerCore}.
\cite{Hochguertel:2008} give a more detailed discussion concerning the internal
FPGA signal processing and the resources consumed in the FFTS cores.

%%% FIGURE :: PFB - MEASURED FREQUENCY RESPONSE OF THREE ADJACENT FREQUENCY BINS (XFFTS-32K)
\begin{figure}
  \centering
  \includegraphics[width=8.9cm]{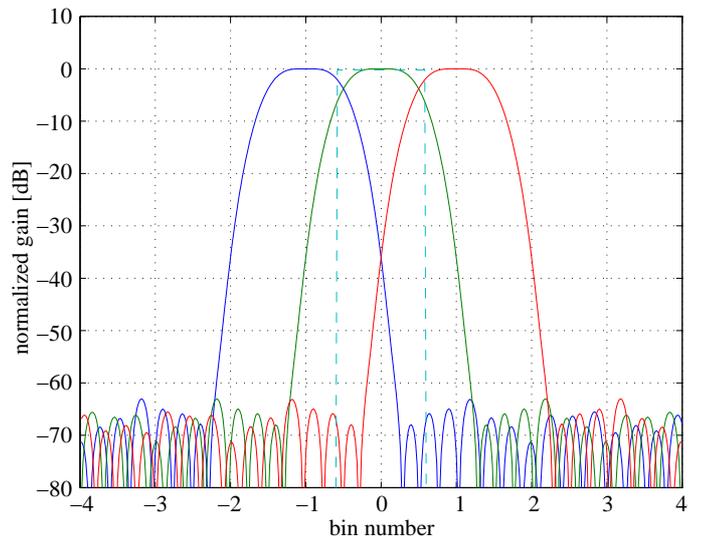}
  \caption{
     Frequency response of the polyphase filter bank signal processing pipeline.
     The diagram shows three adjacent frequency bins in logarithmic scale
     measured with the 32$k$ channel XFFTS.
     The dashed lines illustrate the equivalent noise bandwidth (ENBW) for the
     corresponding spectral bin.
  }
  \label{WOLAfrequencyResponse}
\end{figure}

%%% FIGURE :: PFB - MEASURED SPURIOUS FREE DYNAMIC RANGE (XFFTS-32K)
\begin{figure}
  \centering
   \includegraphics[width=8.9cm]{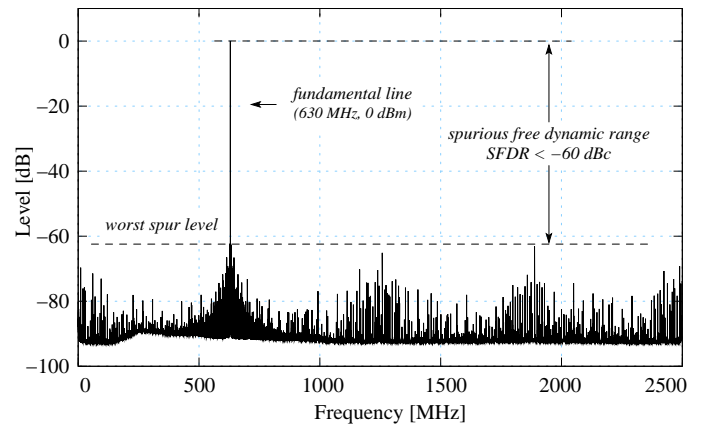}
  \caption{
     Measured spurious free dynamic range (SFDR) of the XFFTS with E2V's
     10-bit 5\,GS/s ADC for a fundamental line injected at 630\,MHz.
     Over the whole 2.5\,GHz bandwidth the SFDR is below $-$60\,dBc,
     perfectly confirming the vendor specifications for this ADC.
  }
  \label{SFDR_XFFTS32K}
\end{figure}

We optimized our polyphase filter coefficients for spectroscopic observations.
The design goal was to find a good trade-off between the frequency resolution and
an optimum use of the limited FPGA resources, particularly the on-chip block-memory.
The equivalent noise bandwidth (ENBW) is generally used to characterize the spectral
resolution. The ENBW is the width of a fictitious rectangular filter such that the
power in that rectangular passband is equal to the integrated response of the actual
filter.
In all our FFTS, the ENBW is adjusted to 1.16\,$\times$ the channel spacing, which
is the total bandwidth of the spectrometer divided by the number of spectral channels.
Figure\,\ref{WOLAfrequencyResponse} illustrates the measured frequency response of
three adjacent frequency bins for a four-tap polyphase filter bank and the corresponding
ENBW for the central bin in logarithmic scale.
The measurements shows a relatively flat passband behavior and a suppression better
than $-$60\,dB for out-of-band signal fractions.
The spurious free dynamic range (SFDR), the intensity ratio of the fundamental signal to the
strongest spurious signal in the power spectrum, was measured to be below $-$60\,dBc across
the 2.5\,GHz bandwidth of the XFFTS (Figure\,\ref{SFDR_XFFTS32K}).
Following the above definition of the ENBW, the spectral resolution of our AFFTS
and XFFTS with default FPGA core is 212\,kHz and 88.5\,kHz, respectively
(Tables \ref{AFFTSpipelineTable}, \ref{XFFTSpipelineTable}).\\

%%% SPECIFICATION AND PERFORMANCE ----------------------------------------------------------------
\section{Specification and performance}
\label{FFTSspecification}

%%% TABLE :: AFFTS HARDWARE
\begin{table}
\caption{Hardware specifications of the {\bf AFFTS} board}
\label{AFFTShardwareTable}
\centering
\begin{tabular}{l r}
\hline\hline
ADC sample rate (default\,/\,max)         & 3\,GS/s / 3.6\,GS/s               \\
Full-scale input power                    & 0\,dBm (630\,mVpp)                \\
Input impedance                           & 50\,$\Omega$                      \\
ADC analog input bandwidth ($-$3\,dB)     & 0.1\,$-$\,3\,GHz                  \\
ADC input resolution (quantisation)       & 8\,Bit                            \\
ADC sampling jitter (10\,$\mu$s record)   & $<$ 500\,fs r.m.s.                \\
ENOB (full bandwidth)                     & 7\,Bits\,@\,748\,MHz              \\
Spurious-free dynamic range (SFDR)        & $>$\,43\,dBc\,@ 1498\,MHz         \\
Differential nonlinearity (typ)           & 0.25\,LSB                         \\
Integral nonlinearity (typ)               & 0.35\,LSB                         \\
FPGA data processing unit                 & XILINX\,Virtex-4\,SX55            \\
GPS/IRIG-B time decoder                   & 1\,kHz, AM\,modulated             \\
GPS/IRIG-B accuracy (decoding)            & $<$\,50\,$\mu$s                   \\
On-board ethernet interface               & IEEE\,802.12, $\leq$\,20\,MBits/s \\
\hline
\end{tabular}
\end{table}
% ------------------------------------------------------------------------------------------

%%% TABLE :: Specifications: AFFTS-processing pipeline
\begin{table}
\caption{Specifications of the {\bf AFFT}-processing pipeline}
\label{AFFTSpipelineTable}
\centering
\begin{tabular}{l r}
\hline\hline
Signal processing / algorithm             & ~~~~~~~polyphase filter bank (PFB)  \\
Processing bandwidth                      & 1.5\,GHz (default)                \\
Spectral channels                         & ~8192 (8$k$)  @ 1.50\,GHz         \\
                                          & 16384 (16$k$) @ 750\,MHz          \\
Channel separation                        & 183\,kHz   @ 1.5\,GHz             \\
Spectral resolution (ENBW)                & 212\,kHz   @ 1.5\,GHz             \\
Internal FPGA signal processing           & 64-bit precision                  \\
On-board integration time                 & max. 5\,s                         \\
Spectral dump time                        & 20\,ms @ 8$k$ channels            \\
Spectroscopic Allan-variance              & $\sim$ 4000\,s                    \\
\hline
\end{tabular}
\end{table}
% ------------------------------------------------------------------------------------------

%%% TABLE :: XFFTS HARDWARE
\begin{table}
\caption{Hardware specifications of the {\bf XFFTS} board}
\label{XFFTShardwareTable}
\centering
\begin{tabular}{l r}
\hline\hline
ADC sample rate				      & max\,5\,GS/s			      \\
Full-scale input power                    & -2\,dBm (500\,mVpp)               \\
Input impedance                           & 50\,$\Omega$                      \\
ADC analog input bandwidth                & 0.1\,$-$\,3.8\,GHz ($-$3\,dB)     \\
ADC input resolution (quantization)       & 10\,Bit                           \\
ADC sampling jitter (10\,$\mu$s record)   & $<$ 500\,fs r.m.s.                \\
ENOB (full bandwidth)		            & 7.2\,Bits\,@\,1.2\,GHz            \\
Spurious-free dynamic range               & $>$\,55 dBc @ 1.2\,GHz            \\
Differential nonlinearity (typ)           & 0.3\,LSB                          \\
Integral nonlinearity (typ)               & 0.8\,LSB                          \\
FPGA data processing unit			& XILINX\,Virtex-6\,LX240T          \\
GPS/IRIG-B time decoder                   & 1\,kHz, AM\,modulated             \\
GPS/IRIG-B accuracy                       & $<$\,50\,$\mu$s (decoding)        \\
On-board ethernet interface               & IEEE\,802.12, $\sim$\,10\,MBytes/s \\
Optional gigabit ethernet interface       & IEEE\,802.3, $\sim$\,90\,MBytes/s \\

\hline
\end{tabular}
\end{table}
% ------------------------------------------------------------------------------------------

%%% TABLE :: Specifications: XFFTS-processing pipeline
\begin{table}
\caption{Specifications of the {\bf XFFT}-processing pipeline}
\label{XFFTSpipelineTable}
\centering
\begin{tabular}{l r}
\hline\hline
Signal processing / algorithm             & ~~~~~~polyphase filter bank (PFB) \\
Processing bandwidth	                  & 2.5\,GHz (default)                \\
Spectral channels			            & 32768 (32$k$) @ 2.5\,GHz	      \\
Channel separation                        & 76.3\,kHz   @ 2.5\,GHz            \\
Spectral resolution (ENBW)                & 88.5\,kHz   @ 2.5\,GHz            \\
Internal FPGA signal processing		& 72-bit precision                  \\
On-board integration time	            & max. 5\,s	                        \\
Spectral dump time			      & 20\,ms @ 32$k$ channels	      \\
Spectroscopic Allan-variance		      & $\sim$ 4000\,s                    \\
\hline
\end{tabular}
\end{table}
% ------------------------------------------------------------------------------------------

%%% FIGURE :: ALLAN STABILITY TEST
\begin{figure}
  \centering
   \includegraphics[width=8.7cm]{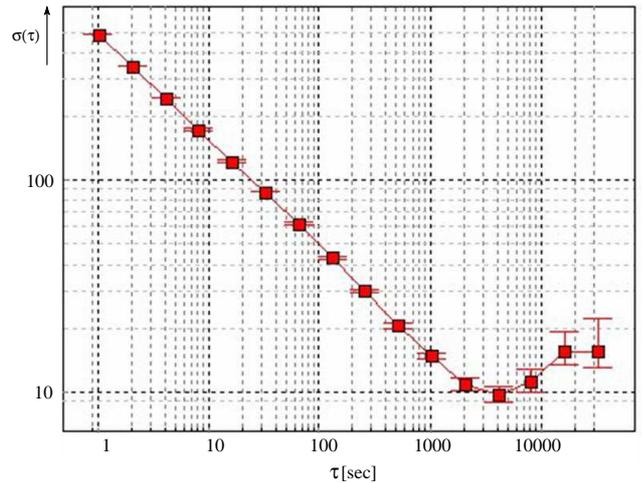}
  \caption{
     Remarkable stability of the FFTS is illustrated in this Allan-variance plot.
     The spectrum of a laboratory noise source was integrated and processed in the AFFTS.
     The spectroscopic variance between two $\sim$\,1\,MHz (ENBW) wide frequency channels,
     separated by 800\,MHz within the band, was determined to be stable on a time scale
     of $\sim$\,4000\,s.
  }
  \label{AllanStabilityTest}
\end{figure}

In Tables \ref{AFFTShardwareTable} and \ref{XFFTShardwareTable} we compile
the hardware specifications and the achieved performances of the AFFTS and
XFFTS boards. The corresponding specifications of the signal processing
pipeline are listed in Tables \ref{AFFTSpipelineTable} and \ref{XFFTSpipelineTable}.

To illustrate the performances of the spectrometer boards, we plot in
Figure\,\ref{AllanStabilityTest} the result of an Allan-variance test,
emphasizing the remarkable stability of the AFFTS on a timescale of
$\sim$\,4000\,s. Similar stability results were demonstrated for the
2.5\,GHz XFFTS.

%%% FFTS INSTALLATIONS ----------------------------------------------------------------
\section{FFT spectrometers in operation}
\label{FFTSinstallations}

The MPIfR FFT spectrometers\footnote{
  Owing to the high demand we outsourced the production and distribution of our
  standard AFFTS and XFFTS. Both spectrometers are manufactured in licence by
  Radiometer Physics GmbH, Germany (http://www.radiometer-physics.de).
} are in use now at several radio telescopes worldwide.
Here we describe the configuration and performance of a few of these installations.

\subsection{AFFTS}
At the APEX telescope an array-FFT spectrometer is in flawless routine operation since
spring 2008 \citep{Guesten:2008}.
In its current configuration it provides a total bandwidth of 32\,$\times$\,1.5 = 48\,GHz
with 256$k$ (32\,$\times$\,8$k$) spectral channels.
By uploading a special FPGA processing core and new ADC synthesizer settings the configuration
can be extended to 58\,GHz (32\,$\times$\,1.8\,GHz) total bandwidth.
This spectrometer is connected to CHAMP$^+$, a 2\,$\times$\,7 pixel heterodyne array
operating in the 350 and 450\,$\mu$m atmospheric windows \citep{Guesten:1998}.

An AFFTS with 24 processing boards has been successfully commissioned at the IRAM 30m-telescope
in July 2011, covering 32\,GHz of total bandwidth with 200\,kHz spectral resolution.
A flexible IF spectrum-slicer, built by IRAM, allows one to observe either four parts of the
IF bands with 200\,kHz resolution and 8\,GHz bandwidth each, or eight parts of the IF bands
with 50\,kHz resolution and 1.82\,GHz bandwidth each.

For two years, an AFFTS with 16 units is in routine operation at the MPIfR
100m Effelsberg telescope.
Owing to the lower operating frequencies of this telescope, the spectrometer covers bandwidths
of 500, 300, and 100\,MHz per board only, but with 16$k$\,spectral channels each by applying a
high-resolution FPGA core.

For GREAT laboratory tests and the Early Science flights, a two-board AFFTS was used
in a 750\,MHz and 1.5\,GHz bandwidth configuration \citep{Heyminck:2012}.

\subsection{XFFTS}
\begin{figure}
  \centering
  \includegraphics[trim = 50mm 0mm 0mm 0mm, clip, angle=270,width=8.9cm]{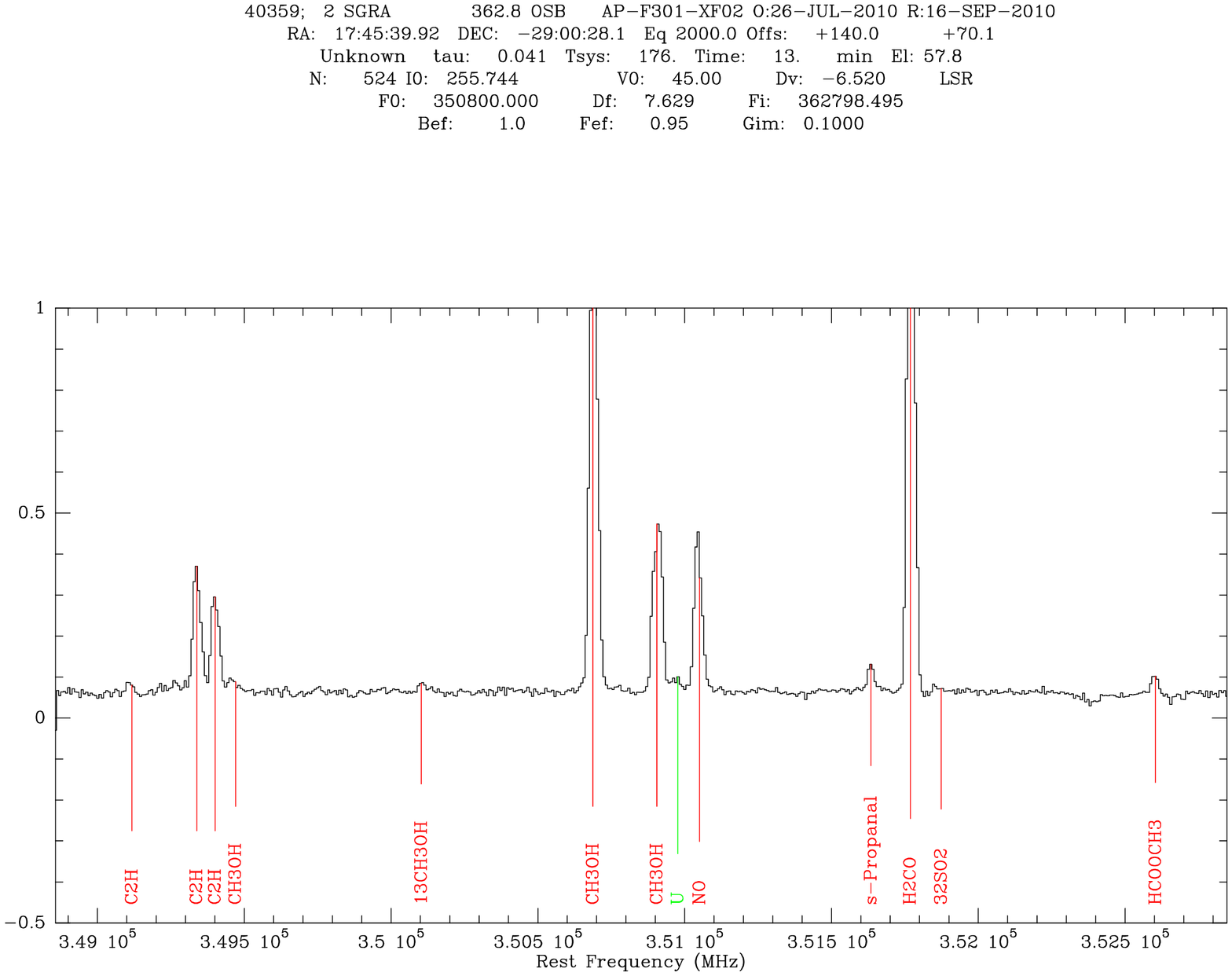}
  \caption{
      Two 2.5\,GHz XFFT boards combined, with 500\,MHz overlap, to create a 4\,GHz-wide spectrometer.
      The sample spectrum, part of a Galactic Center line survey (G{\"u}sten, priv comm.),
      shows the upper sideband of the sideband separating APEX/FLASH$^+$ receiver.
  }
  \label{XFFTSspectrum}
\end{figure}

A first version of the 2.5\,GHz XFFTS has been successfully commissioned in 2010
at the APEX telescope. In combination with a flexible IF signal processor,
four XFFTS boards cover a bandwidth of 2\,$\times$\,4\,GHz bandwidth with a
spectral resolution (ENBW) of 88.5\,kHz and 500\,MHz overlap in the band center.
This setup allows one to observe the upper and lower sideband of the re-modeled
sideband separating FLASH$^+$ receiver at the same time (Fig.\,\ref{XFFTSspectrum}).

Since July 2011, two XFFTS boards -- in standard configuration -- are in regular
operation with GREAT, providing 2.5\,GHz bandwidth with a spectral resolution of
88.5\,kHz (ENBW).
Even under the harsh environmental conditions at facilities such as APEX or SOFIA,
the FFT spectrometers have proven to be extremely reliable and robust.

Finally, at the 100m Effelsberg telescope, a two-unit XFFTS has been successfully
commissioned at the end of 2011, with remote switchable FPGA configurations for
high-resolution spectroscopy at 2\,GHz, 500\,MHz, and 100\,MHz bandwidths.

\label{FFTSobservation}
%%% FIGURE :: XFFTS-SPECTRUM

%%% CONCLUSION AND OUTLOOK ------------------------------------------------------------------
\section{Conclusions and outlook}
\label{FFTSoutlook}
We briefly summarize the advantages of our high-resolution wide-band Fast Fourier Transform spectrometers:
\begin{itemize}
  \item
  FFT spectrometers provide high instantaneous bandwidth (1.5\,GHz, 2.5\,GHz)
  with many thousands of frequency channels, thus offering wide-band observations
  with high spectral resolution.

  \item
  The implemented polyphase filter bank algorithm provides a nearly loss-free
  time-to-frequency transformation with significantly reduced frequency scallop,
  less noise bandwidth expansion, and faster sidelobe fall-off.

  \item
  FFTS have been demonstrated to operate extremely stable due to exclusive digital signal processing.
  Allan-variance stability times of several 1000 seconds have been achieved routinely.

  \item
  FFTS require only small volume, mass, and power budgets -- ideal for operation at high-altitude
  observatories (e.g., APEX at 5100-m) as well as on airborne observatories such as SOFIA.

  \item
  Production costs are low compared to traditional spectrometers because only commercial
  components and industrial manufacturing are used.

  \item
  The superior performance, high sensitivity and reliability of our FFT spectrometers
  have been demonstrated at many telescopes word-wide.
\end{itemize}

The announcements of new ADCs with even higher sample rates ($f_s$ $\geq$ 10\,GS/s)
and wider analog input bandwidth, together with the still increasing processing
capability of future FPGA chips (e.g., the Xilinx Virtex-7 series), make it very likely
that FFT spectrometers can be extended to even broader bandwidth with adequate numbers
of spectral channels in the near future.
Our next FFTS development will be the design of a spectrometer with instantaneous bandwidth
$\geq$\,4\,GHz and up to 128$k$ spectral channels, aiming at operational readiness in time for
the commissioning of our upGREAT detector array in
2013/14{\footnote{The projected development time is based on a time-to-science figure-of-merit
  of $\sim$18 months that we have demonstrated during  previous implementations of new DSP
  opportunities. Timely injection of new science opportunities is essential for the success of
  a PI instrument like GREAT, flying cutting-edge technologies on-board an expensive mission.
  For a technology that still advances according to Moore's growth law our highly specialized
  approach ensures probably better, faster realization than designs based on hardware building
  blocks from a DSP developer platform like CASPER 
  (see, e.g., \citealt{Parsons:2005, Parsons:2009} about their philosophy to support broader
   community applications with open hardware architectures and open signal processing libraries).}
}. \\
In current FFTS applications an IF processor is needed for down-mixing the receiver signals
to baseband.
However, these analog processors are time-consuming to build, not calibration- and aging-free,
and cost-intensive.
The notification of a new class of track-and-hold amplifiers, operating at GHz frequencies,
will allow direct sampling of the intermediate frequency of current and future heterodyne
receivers by bandpass-sampling techniques, with much reduced complexity and reduced costs.

%%% BIBLIOGRAPHY ------------------------------------------------------------------
\bibliographystyle{aa}

\bibliography{xffts}

\end{document}